\documentclass[%
      prb, 
      twocolumn,
      showkeys,
      superscriptaddress,
      showpacs]{revtex4-2}
      
\usepackage[utf8]{inputenc}
\usepackage[T1]{fontenc}

\usepackage{graphicx}
\graphicspath{ {figures/} }
\usepackage{amsmath, amsfonts, amssymb, mathtools}	
\usepackage{xcolor}

\begin{document}

\title{Strain-induced exciton mobility in layered WS$_2$ from first principles}

\author{Amir Kleiner}
\email[Corresponding author: ]{amir.kleiner@weizmann.ac.il}
\affiliation{Department of Molecular Chemistry and Materials Science, Weizmann Institute of Science, Rehovot 7610001, Israel}

\author{Sivan Refaely-Abramson}
\email[Corresponding author: ]{sivan.refaely-abramson@weizmann.ac.il}
\affiliation{Department of Molecular Chemistry and Materials Science, Weizmann Institute of Science, Rehovot 7610001, Israel}



\date{\today}

\begin{abstract}
Exciton mobility in two-dimensional semiconductors is a key ingredient in materials-based design of optoelectronic functionalities. Monolayer transition metal dichalcogenides (TMDs) set a good test case, with tightly bound excitons and designable flexibility that offer an ideal platform for realizing strain effects on exciton energy transfer. Here, we present an ab initio study to construct strain-induced exciton energy profiles and model exciton dynamics on top of these potential surfaces. We focus on inhomogeneously-strained monolayer WS$_2$, combining excitonic band structures derived from many-body perturbation theory for a large variety of strain profiles and calculate the change in mobility characteristics using a semiclassical ballistic transport model. We connect a wealth of strain patterns to exciton drift, diffusion, and confinement. Our results point to strain-induced regimes of super-ballistic propagation and an anomalous effective diffusion, governed entirely by the strain landscape. Our results provide structure-specific understanding of ballistic strain-tunable exciton behavior, offering design principles for engineering exciton dynamics in two-dimensional materials.

\end{abstract}

\maketitle

Transition metal dichalcogenides (TMDs) are a class of wide band-gap semiconductors whose optoelectronic properties strongly depend on their layered structure. While bulk TMDs exhibit an indirect gap, their monolayer (ML) forms possess a direct band gap, with both the conduction and valence band extrema shifting between the two limits \cite{mak_atomically_2010, yun_thickness_2012}. Together with the strongly bound excitons formed upon light excitation, these properties led to the emergence of monolayer TMDs as a versatile platform for engineering quantum phenomena, with structural tunability via chalcogen vacancies \cite{refaely-abramson_defect-induced_2018, amit_tunable_2022, hernangomez-perez_reduced_2023}, heterostructure design \cite{geim_van_2013, latini_excitons_2015, zeng_novel_2018}, and control over twist angles and moiré superlattices \cite{kleiner_designable_2024, kundu_moire_2022, merkl_ultrafast_2019}.
One of the common approaches for tuning the electronic and excitonic properties of TMD monolayers is the application of in-plane strain, which provides a continuous and controllable means of modulating both the electronic and excitonic behavior. Monolayer TMDs can sustain strains exceeding $\sim11 \%$ without mechanical failure or phase transitions \cite{feng_strain-engineered_2012, ghorbani-asl_strain-dependent_2013}, and their band structures exhibit pronounced sensitivity to such deformations. 

Applied strain shifts both the positions and curvatures of band extrema \cite{feng_strain-engineered_2012, ghorbani-asl_strain-dependent_2013, shi_quasiparticle_2013, thompson_anisotropic_2022}, with effects that vary across the Brillouin zone and enable smooth transitions between direct and indirect band gaps \cite{shi_quasiparticle_2013, desai_strain-induced_2014}. The excitonic band structure is similarly strain-sensitive, showing tunable dispersion and energy shifts under deformation \cite{deilmann_finite-momentum_2019}. While this general strain response is consistent across various semiconducting TMD monolayers \cite{guzman_role_2014, zollner_strain-tunable_2019} and many multilayer systems \cite{sharma_strain_2014, wang_strain-induced_2020}, the precise strain dependence and the critical strain for gap transitions vary with composition and thickness \cite{wang_strain-induced_2015, defo_strain_2016, khatibi_impact_2018}.
Spatially varying strain in monolayer TMDs generates inhomogeneous or non-uniform strain profiles that can actively drive exciton dynamics in the material \cite{feng_strain-engineered_2012, cordovilla_leon_exciton_2018, moon_dynamic_2020-1, datta_spatiotemporally_2022, kim_strained_2024}, enabling exciton trapping and guiding via strain-engineered potentials \cite{castellanos-gomez_local_2013, wang_strain-induced_2020, dirnberger_quasi-1d_2021, li_enhanced_2023}. From a theory perspective, these strain-induced effects are typically treated as exciton drift driven by energy gradients \cite{feng_strain-engineered_2012, moon_dynamic_2020-1}, as well as strain-induced diffusion \cite{cordovilla_leon_exciton_2018}. 

Exciton transport in such systems has typically been modeled using phenomenological drift-diffusion equations, where position- and time-dependent mobility and diffusion coefficients are fitted to experimental or empirical trends \cite{cordovilla_leon_exciton_2018, harats_dynamics_2020}. Drift is commonly attributed to the strain-induced spatial variation of exciton energies—driving excitons toward regions of lower energy (higher strain) \cite{feng_strain-engineered_2012, cordovilla_leon_exciton_2018}.
Diffusion behavior, in contrast, has drawn considerable attention due to reports of anomalous \cite{rosati_strain-dependent_2020, uddin_neutral_2020, wagner_nonclassical_2021} and even so-called “negative” diffusion regimes \cite{rosati_negative_2019}. Yet most theoretical treatments rely on simplified models that either neglect the many-body quasiparticle nature of the excitons \cite{khatibi_impact_2018, rosati_dark_2021, an_strain_2023}, or use first principles results to parametrize otherwise phenomenological transport models \cite{rosati_strain-dependent_2020, chand_visualization_2022, hernandez_lopez_strain_2022}. In particular, excitonic dispersion is often approximated using strain-independent effective masses, and energies are typically evaluated at high-symmetry k-points \cite{rosati_strain-dependent_2020, wagner_nonclassical_2021}. While such approaches can capture complex many-body processes such as intervalley scattering \cite{rosati_dark_2021, an_strain_2023} or trion formation \cite{harats_dynamics_2020, lee_drift-dominant_2022}, they often neglect the full momentum-space structure of the exciton under strain, limiting their ability to track the structure-specific mechanisms driving exciton transport in strained 2D materials.

In this work, we present a first–principles–based framework to study exciton mobility as a function of inhomogeneous strain. We demonstrate our approach for the case of WS$_2$, a prototypical monolayer TMD. Using many-body perturbation theory within the GW and Bethe Salpeter equation (GW-BSE) approach, we compute both real- and momentum-space potential exciton energy surfaces. We explore characteristic properties of the exciton motion under a large variety of strain landscapes, isolating the impact of band curvature and energy gradients on drift, confinement, and diffusion, to detect how strain governs distinct transport regimes. Our approach is generally applicable to quasiparticles in systems of arbitrary dimensionality where the band structure evolves continuously with strain, offering a general pathway for understanding the design principles for strain-guided excitonic and quantum transport in next-generation optoelectronic and excitonic materials and devices.

\begin{figure}
    \centering
    \includegraphics[width=1.\linewidth]{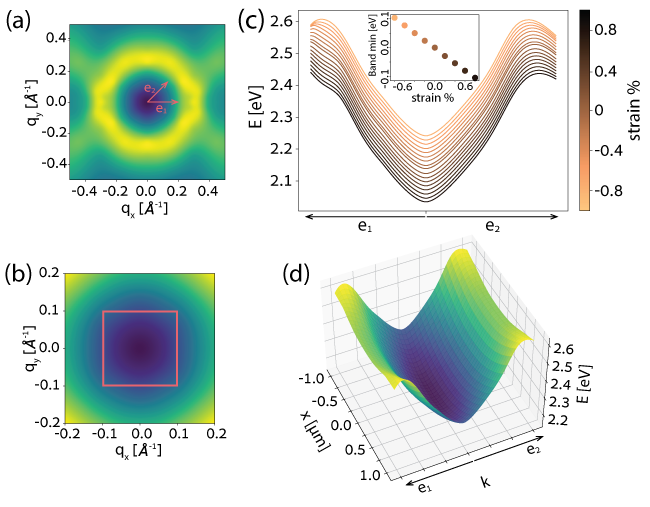}
    \caption{
    (a) First bright excitonic band centered around the $\Gamma$ center-of-mass momentum with paths utilized for line-plot marked as $e_1$ and $e_2$. (b) Close-up of the parabolic region of the band, with the region utilized for the dynamics simulations marked by a red square.
    (c) First bright excitonic band along the $e_1$ and $e_2$ directions, plotted for various strains.
    (inset) position of the band minimum for various strains. The exciton band minimum exhibits an inverse proportionality for the range of strains utilized in this study.
    (d) Illustration of a 2D slice of the full potential, with a concave parabolic strain profile along one direction in position space, and a path along $e_1$ and $e_2$ in momentum space, representing the (1,1) and (1,-1) directions in crystal axes.
    }
    \label{fig:Fig1}
\end{figure}

\section*{Strained and position-dependent band structures}

Experimentally, strain profiles can be induced in materials by various methods - e.g. by applying direct forces through an AFM cantilever \cite{feng_strain-engineered_2012, harats_dynamics_2020, koo_tip-induced_2021} or multi-point bending apparatus \cite{mennel_second_2018}, pressurizing an airtight cavity with a monolayer TMD as a membrane \cite{kovalchuk_non-uniform_2022}, patterning and imprinting \cite{bensmann_nanoimprint_2022, chand_visualization_2022}, wrinkling \cite{dhakal_local_2017, koo_tip-induced_2021} or electrostatic forces \cite{hernandez_lopez_strain_2022}.
Each method results in drastically different, ranging from simple linear to very complex non-analytical strain profiles.
Here we restrict our study to simple polynomial strain profiles, aiming to elucidate the physical principles behind the strain-induced dynamics, and their dependence on the shape of the strain profile.

We construct a position- and momentum-dependent exciton potential for a series of deformed monolayer WS$_2$ systems, each subjected to uniform strain by modifying the unit cell dimensions and internal atomic coordinates accordingly:
Tensile strain is applied by enlarging the unit cell, while compressive strain is applied by shrinking it.
In this study we focus on uniform strains as an illustrative example.
The atomic geometry of the unit cells remains unchanged after their modification. 
For each strained configuration, we perform Density Functional Theory (DFT) \cite{perdew_generalized_1996,giannozzi_quantum_2009,giannozzi_advanced_2017,giannozzi_quantum_2020,monkhorst_special_1976} calculations of the electronic states and energies, followed by a calculation of the quasiparticle (QP) corrections within many-body perturbation theory within the G\textsubscript{0}W\textsubscript{0} approximation \cite{hybertsen_electron_1986,deslippe_berkeleygw_2012,del_ben_large-scale_2019,wu_spin-orbit_2020, da_jornada_nonuniform_2017}. 
The exciton band structure is then obtained by solving the Bethe-Salpeter equation (BSE)\cite{rohlfing_electron-hole_2000, yang_excitonic_2009, qiu_nonanalyticity_2015}, yielding the momentum-resolved exciton energies required for our dynamics model.
Full computational details are given in the SI~\cite{supp}.

The atomic arrangement in TMD monolayers has a hexagonal symmetry, manifested in the excitonic band structure as shown in Figure \ref{fig:Fig1} (a).
Near the $\Gamma$-point (zero center-of-mass momentum), the first bright exciton band exhibits a parabolic dispersion \cite{qiu_signatures_2021}, shown in figure \ref{fig:Fig1} (b).
Applying strain to the monolayer modulates both the curvature of the exciton dispersion and the absolute energy of the band minimum.
We calculate induced strains in the range of $-0.8 \%$ to $+0.8 \%$ applied to the whole unit cell. Such strains have a small effect on the effective masses close to $\Gamma$, though the effect gets more pronounced further away from $\Gamma$. At this range the shift in the energy of the band edge is linear with strain, with a magnitude of $129.6$ meV\/$\%$, in close agreement with previously measured and computed values \cite{blundo_strain-induced_2022,kourmoulakis_biaxial_2023}.
These trends are showcased in Figure \ref{fig:Fig1} (c).
When an inhomogeneous strain profile is applied to a large layer, the shape of each unit cell is altered. However, provided that the strain gradient is small compared to the unit cell size, locally the changes are negligible, and a local band structure approximation remains valid.
This assumption enables us to construct a position-dependent band structure by interpolating the strained band structures across a continuous strain field.
For a 2D layer, this results in a 4D potential $E(\vec{r}, \vec{q})$ that explicitly depends on both the two position-space coordinates $\vec{r}$ and the two crystal momentum-space coordinates $\vec{q}$.
Figure \ref{fig:Fig1} (d) shows a 2D slice through one such 4D potential surface, combining a 1D concave parabolic strain profile along position space with a momentum-space cut through the Brillouin zone. The resulting energy landscape reflects the anisotropic and nonlinear coupling between strain and exciton dynamics. Further details of the interpolation procedure are provided in the SI~\cite{supp}.
This interpolated potential allows us to investigate the exciton time evolution via a simplified model, in which the strain-induced band structure variations translate into spatial and ballistic movement of an exciton wavepacket.

\section*{Exciton wavepacket dynamics}
While the simulations are not meant to replicate experimental conditions quantitatively, they provide valuable qualitative insight into the ballistic transport regimes relevant for strain-engineered exciton dynamics. We consider the case where a large enough number of excitons is generated, such that we can employ Ehrenfest's theorem to keep track of the exciton wavepacket center-of-mass (COM) position $\vec{r}$ and momentum $\vec{q}$.
We thus model the dynamics of a macroscopic exciton distribution $n(\vec{r},\vec{q})$, initialized by an optical excitation centered around a point $\vec{r}_0$ with a width consistent with that of a diffraction-limited excitation source, and characterized by a spread in momentum around the center-of-mass $\vec{q}=0$ consistent with the light cone of the excitation source.
The ballistic time evolution of this distribution on the position- and momentum-dependent potential energy surface is governed by the following general equation of motion (EOM):
\begin{equation}
\begin{split}   
    \frac{\partial n_X(\vec{r},\vec{q},t)}{\partial t} & =  \\
    & \nabla_{\vec{q}} n_X(\vec{r},\vec{q},t) \cdot \nabla_{\vec{r}} E(\vec{r},\vec{q})  \\
    - & \nabla_{\vec{r}} n_X(\vec{r},\vec{q},t) \cdot \nabla_{\vec{q}} E(\vec{r},\vec{q}) 
    \label{eq:EOM}
\end{split}
\end{equation}

Equation \ref{eq:EOM} reflects a semiclassical approximation to the exciton dynamics, appropriate for modeling the evolution of macroscopic exciton ensembles, as often probed in experimental studies and envisioned in excitonic device applications. We note that in the non-ballistic case, an additional scattering term $K_{ext}$ must be added that incorporates all external interactions, such as exciton-phonon and exciton-exciton scattering, interband crossings, and spontaneous radiative decay (see SI~\cite{supp} for further details on the chosen EOM). 
Due to the computational complexity, here we neglect these effects and focus on the ballistic regime, searching for fundamental relations between the strain-induced excitonic potential surface and the exciton mobility character.
While this choice limits the validity of our results on larger time-scales, it allows for a fully ab initio calculation of multiple strain profiles and a comprehensive understanding of these structural effects. 
The position and momentum-dependent potential does not, in general, correspond to a conservative force. The variation of the strain with position introduces effective forces that redistribute the exciton ensemble in both position and momentum space.
This not only drives drift in position space but also evolves the momentum distribution, enabling us to track how exciton dynamics emerge from strain-induced band structure modulations.
The total energy of the distribution is conserved under the ballistic limit of the EOM. This results in a confinement of the momentum-space distribution for small strain values such as those studied in this work. 

The potential energy surface was evaluated on a square real-space grid spanning $2 \mu \textnormal{m} \times 2 \mu \textnormal{m}$ with $20 \textnormal{nm}$ resolution, and a momentum-space grid from $-0.1$ \AA$^{-1}$ to $0.1$ \AA$^{-1}$ with $2\times 10^{-3}$ \AA$^{-1}$ resolution.
Strain profiles were chosen to span the computed range of uniform strain values ($\pm 0.8\%$) without extrapolation.
The profiles studied have the general shape
\begin{equation}
    s(x,y)=a_0+a_1x+a_2x^2+a_3x^3+b_1y+b_2y^2+b_3y^3
    \label{eq:general_strain}
\end{equation}
where we can define separately \emph{linear}, parabolic \emph{concave}, parabolic \emph{convex}, or \emph{cubic} profile for each position coordinate.
An initial exciton distribution was then generated in various positions on the surfaces, with varying widths in the range of $30-50$ nm, which corresponds to a photon beam of width $100-170$ nm, which is about 2 times smaller than the diffraction limit for a laser in resonance with the unstrained A exciton.
In momentum space, the distribution was initialized around zero-momentum, with a width varying in the range of $3\times10^{-2}$ to $2.5\times10^{-3}$, which corresponds to a distribution between an order of magnitude and a factor of 2 larger than the cone of light for a laser in resonance with the unstrained A exciton.
Both widths have been chosen due to numerical stability considerations. 
The time evolution of the distribution is evaluated using a 4th-order Runge-Kutta method, with 5th-order error estimations (RK45), while the static right-hand side of the equation is computed explicitly after each time step.
We run the simulation for a total of $\sim 1.5$ ps. The temporal resolution is automatically determined by the RK45 solver, with a typical time step of 1-10 fs.

\section*{Strain effects on exciton mobility}
We generated 11 distinct inhomogeneous strain profiles by activating individual polynomial components in each spatial dimension of Eq.~\ref{eq:general_strain}. For each profile, we initialized 18 exciton distributions with varying spatial positions and momentum-space widths, allowing systematic exploration of how both strain geometry and initial conditions shape exciton dynamics.
\begin{figure}
    \centering
    \includegraphics[width=1.\linewidth]{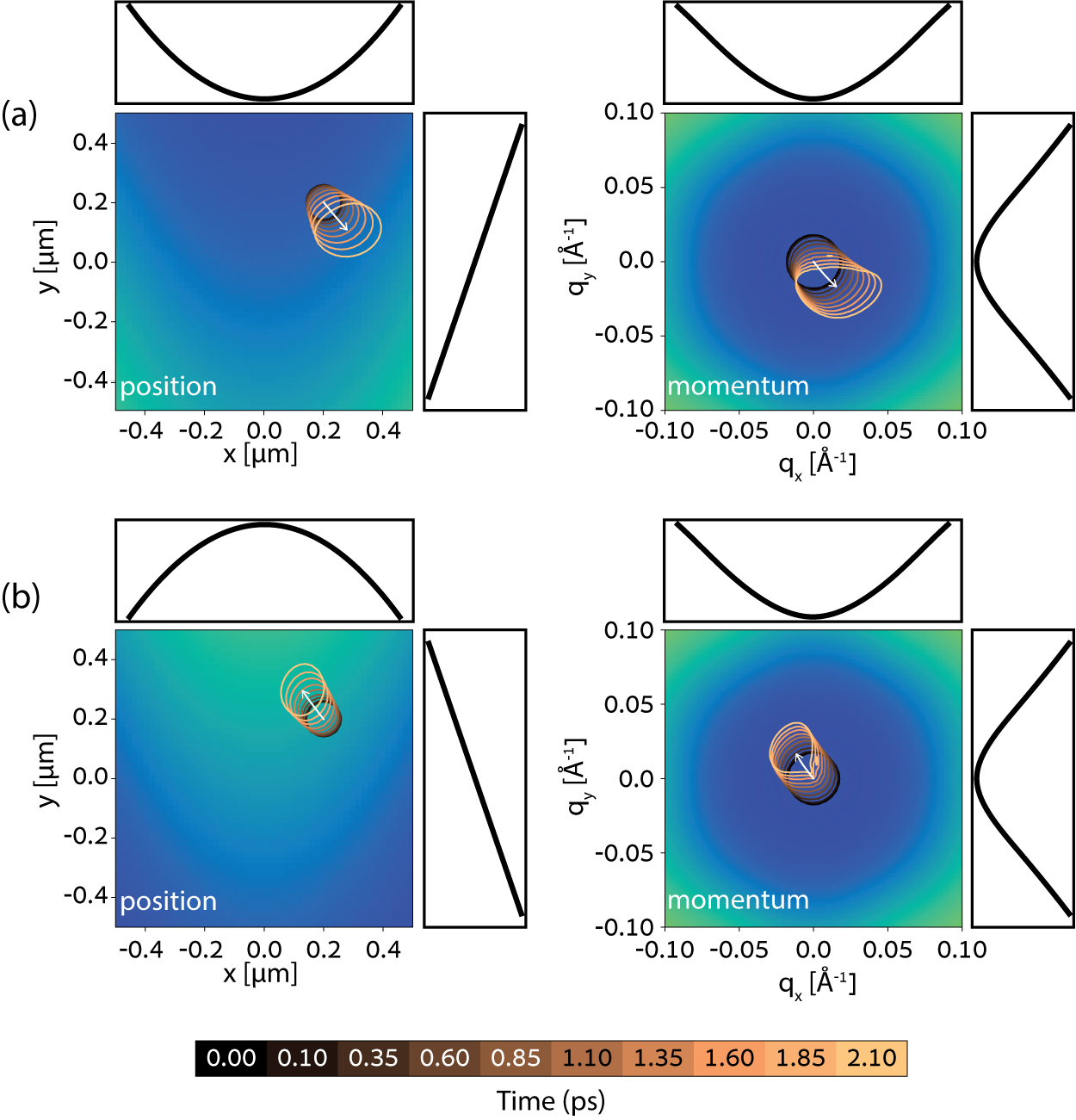}
    \caption{Time evolution of the exciton distributions, expressed as the contour at 50\% height for multiple time-steps and the propagation of the COM as a white arrow in position space (left) and momentum space (right) for (a) a parabolic convex in X and linear ascending in Y and (b) a parabolic concave in X and linear descending in Y strain profiles.
    Position-space backgrounds illustrate the appropriate strain profile. Low strain values are in blue, and high values are in green. As a guide to the eye, the projections along the axes are illustrated along the appropriate axes.
    Momentum-space backgrounds illustrate the average excitonic band. Low energy values are in blue, and high values are in green. As a guide to the eye, the projections along the axes are illustrated along the appropriate axes.
}
    \label{fig:Fig2}
\end{figure}
Figure \ref{fig:Fig2} illustrates representative cases of exciton propagation under combined strain profiles: convex or concave variation in the X-direction, coupled with linearly increasing or decreasing strain in the Y-direction.
The evolution of the exciton distribution in both position and momentum space is shown as 50\% height contours at multiple time steps, with white arrows tracking the distributions' COM trajectory.
The position-space background schematically illustrates the strain profile present in the system, while the momentum-space background schematically illustrates the average excitonic band in the system. In both cases, low-strain/energy values are represented in blue, and high-strain/energy values are represented in green. 
More detailed snapshots of the dynamics are presented in the SI~\cite{supp}.

The coupling between real- and momentum-space components in the strain-induced potential results in complex propagation behavior, including significant distortion of the distribution in both domains.
The COM of the distribution is used to track and analyze the drift elements of the dynamics. 
Consistent with experimental observations \cite{cordovilla_leon_exciton_2018, dirnberger_quasi-1d_2021}, the exciton distributions exhibit directional drift toward regions of higher strain.
The total drift depends largely on the type of strain profile and initial conditions. The maximal drift distance observed throughout the simulation is $0.18 \mu$m, which results in a maximal average drift velocity of $1.20\times 10^{2} \mu\textnormal{m}/\textnormal{ns}$. 
The resulting drift velocity of approximately $120 \mu\textnormal{m}/\textnormal{ns}$ is $\sim3$ orders of magnitude larger than experimental values \cite{cordovilla_leon_exciton_2018}, reflecting the idealized ballistic regime and suggesting a mean free time on the order of $1.5$ fs.

\begin{figure}
    \centering
    \includegraphics[width=1.\linewidth]{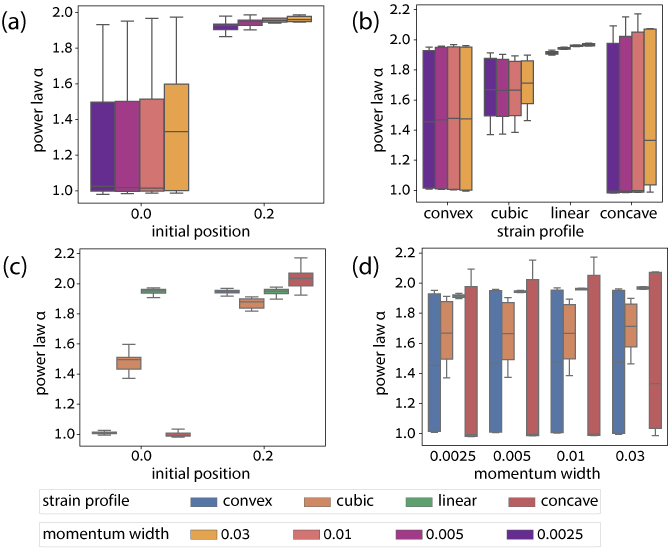}
    \caption{Box plot of the strain-induced mobilities as reflected in the power law power law $\vec{r}(t)\propto t^\alpha$, resulting under (a) various initial positions with solutions grouped by initial width of momentum distribution, (b) various strain profiles with solutions grouped by initial width of momentum distribution, (c) various initial positions with solutions groups by shape of strain profile, (d) various initial width of momentum distribution with solutions groups by shape of strain profile. 
    A purely ballistic propagation should result in linear mobility.
    Most strain profiles yield sub-ballistic transport behavior ($\alpha < 1$), but concave parabolic profiles in particular lead to ballistic or even super-ballistic regimes ($\alpha \approx 1 - 2$).
    The researched strain profiles are either \emph{linear}, parabolic \emph{concave}, parabolic \emph{convex}, or \emph{cubic} with respect to a position coordinate, as per equation \ref{eq:general_strain}. }
    \label{fig:Fig3}
\end{figure}

We define the strain-induced mobility as the response of the velocity to the applied force $\eta_\varepsilon(t) = \vec{v}_{drift}(t)\cdot \nabla_{\vec{r}}\varepsilon(t)$. 
This can be generally expressed as a simple power law $\vec{r}(t)\propto t^\alpha$, where the value of $\alpha$ determines the regime of the motion. $\alpha=1$ corresponds to a, constant-motion, purely ballistic propagation, values of $\alpha<1$ correspond to sub-ballistic drift, and values of $\alpha>1$ correspond to super-ballistic drift. A special case of notice is $\alpha=2$, which corresponds to propagation under a constant force.
Figure \ref{fig:Fig3} shows the power laws resulting from the time dependence of the mobility.
To understand the effect of the various initial states and strain profiles on the mobility, the results are shown using box plots, and the different simulations were aggregated according to combinations of pairs of such conditions.
Further details on this processing are provided in the SI~\cite{supp}.
Although some simulations yield a slope of exactly 1 or 2, most fall between these values, with very few having a value lower than 1, and very few have values higher than 2.
These results suggest that, predominantly, the exciton drift exhibits super-ballistic characteristics with a driving force that decreases with time.
This behavior arises from the shape of the dispersion relations and their dependence on strain, which results in the addition of both accelerating and damping effective forces.
Notably, although the shape of the strain profile and the initial position on the profile have a large effect on the drift regime, the initial momentum width has a negligible effect.

\begin{figure}
    \centering
    \includegraphics[width=1.\linewidth]{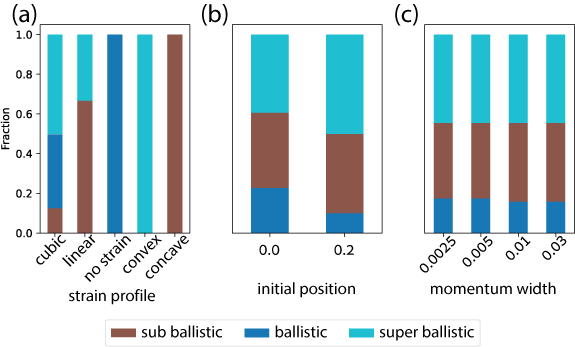}
    \caption{Behavior of the time evolution of position-space width under (a) various strain profile shapes, (b) various initial positions, and (c) various initial momentum widths.
    Without strain, the time-evolution of the position-space width behaves as a ballistic expansion. Concave strain profiles always result in a sub-ballistic expansion, which is related to negative diffusion regimes. Convex strain profiles always result in super-ballistic expansion. Linear strain profiles can induce either sub- or super-ballistic expansions depending on initial conditions, while cubic profiles can exhibit all three expansion behaviors.
    Both the sub-ballistic and super-ballistic regimes correspond to anomalous diffusion regimes when interparticle interactions are considered.}
    \label{fig:Fig4}
\end{figure}

A particularly intriguing aspect of exciton transport in strained TMDs is the emergence of anomalous diffusion regimes, including reports of so-called "negative" diffusion, where exciton distributions appear to contract rather than broaden over time. While proper diffusion typically arises from interparticle interactions, ballistic broadening contributes to the overall spatial expansion and can serve as a proxy for interpreting such anomalous behaviors. In anisotropic strain landscapes, exciton distributions often become distorted, prompting us to analyze the time evolution of distribution widths projected along orthogonal directions, where the profiles remain approximately Gaussian throughout propagation.
In the absence of strain, exciton broadening follows the expected ballistic trend, where the square of the spatial width scales linearly with the square of time for all initial momentum-space widths. The slope of this relationship depends on the initial spread in momentum space, consistent with semiclassical expectations. Under nonzero strain, however, the broadening deviates from this ideal. To classify these deviations, we introduce a normalized width by comparing each strained system to its corresponding unstrained counterpart. A normalized width greater than one indicates super-ballistic broadening; values below one indicate sub-ballistic behavior.

Figure~\ref{fig:Fig4} presents the frequency of these regimes across different strain profiles, initial exciton positions, and momentum-space widths. The strain profile geometry emerges as the dominant factor shaping the broadening behavior, while the initial momentum width plays a negligible role. Parabolic strain profiles yield consistent behavior: convex profiles produce super-ballistic broadening, while concave profiles lead to sub-ballistic broadening. Linear strain profiles tend to exhibit super-ballistic broadening unless coupled with nonlinearity in an orthogonal direction, which then suppresses the expansion. Cubic profiles display position-dependent behavior: excitons initialized in convex regions broaden super-ballistically, while those in concave regions are sub-ballistic. At inflection points, broadening is approximately ballistic.
Interestingly, combinations of sub-ballistic broadening and suppressed drift in the same direction give rise to negative broadening. These results can provide a possible explanation for the experimentally observed anomalous and negative diffusion, and demonstrate how the geometry of the strain landscape can be harnessed to control exciton transport in 2D materials.

\section*{Conclusions}

To conclude, we presented a first-principles-based framework for modeling the ballistic time evolution of exciton distributions in inhomogeneously strained semiconductors, applied to strained monolayer WS\textsubscript{2}. Exploring idealized polynomial strain profiles, our results capture the qualitative impact of band curvature and energy gradients on exciton drift and diffusion patterns, in the absence of scattering and relaxation mechanisms.
The model qualitatively reproduces the experimentally observed behaviors of drift toward strain maxima, anisotropic broadening, and negative-like diffusion. It also provides a physical basis for estimating the ballistic mean free time in idealized regimes.
These results offer fundamental insight into how strain modulates exciton transport in 2D semiconductors and establish a foundation for ab initio-informed design principles in optoelectronic devices, where strain landscapes are engineered to control quasiparticle motion.

\noindent \textbf{Acknowledgments:}
We thank Moshe Harats, Lev Melnikovski, Diana Qiu, and Galit Cohen for valuable discussions. Computational resources were provided by the ChemFarm local cluster at the Weizmann Institute of Science and the Max Planck Computing and Data Facility cluster. This research was supported by a European Research Council (ERC) Starting Grant (No. 101041159).


\section*{Data availability}
The raw exciton band structure data generated and utilized in this study is available as part of the code base at the public github repository \url{https://github.com/amirkl91/strain_dynamics} with \url{https://doi.org/10.5281/zenodo.15855463}.
\section*{Code availability}
All code utilized to generate potential surfaces and solve dynamics model is available from the public github repository \url{https://github.com/amirkl91/strain_dynamics} with \url{https://doi.org/10.5281/zenodo.15855463}.

\bibliographystyle{naturemag}
\bibliography{biblio}

\end{document}